\documentclass[aps, prb, letterpaper, 10pt, twocolumn, superscriptaddress]{revtex4-2}

\usepackage[colorlinks=true,allcolors=blue!80!black]{hyperref}
\usepackage{nameref} 
\usepackage{comment}
\usepackage[utf8]{inputenc}
\usepackage{amsmath}
\usepackage[caption = false]{subfig}
\usepackage{graphicx,epstopdf}
\usepackage{blindtext}
\usepackage{lipsum}
\usepackage{amsfonts}
\usepackage{bbm}
\usepackage{amssymb}
\usepackage{enumerate}
\usepackage{color}
\usepackage{latexsym}
\usepackage{physics}
\usepackage{times,txfonts}
\usepackage{xcolor}
\usepackage{epsfig}
\usepackage{bm}
\usepackage{tabularx}
\usepackage{multirow}
\usepackage{mathtools}
\usepackage{xfrac}
\usepackage{physics}
\usepackage{enumerate}
\usepackage{braket}
\usepackage{graphicx}
\usepackage[T1]{fontenc} 

\begin{document}
\title{Interfering-or-not-interfering quantum key distribution with advantage distillation}

\author{Seyede Zahra Zarei}
\email{Rahinzarei@gmail.com}
\affiliation{Department of Physics, Isfahan University of Technology, Isfahan 84156-83111, Iran}
\author{Fatemeh Tarighi Tabesh}
\email{f.tabesh@ipm.ir}
\affiliation{School of Physics, Institute for Research in Fundamental Sciences (IPM), Tehran 19395-5531, Iran}
\author{Mehdi Abdi}
\email{mehabdi@gmail.com}
\affiliation{Wilczek Quantum Center, School of Physics and Astronomy, Shanghai Jiao Tong University, Shanghai 200240, China}

\date{\today}
\begin{abstract}
Interfering-or-not-interfering quantum key distribution (INI-QKD) is an innovative protocol whose performance surpasses existing twin-field protocol variants.
In this study, we introduce an additional step of advantage distillation (AD) after the quantum communication phase to further enhance its performance.
Through the AD the raw key is partitioned into small blocks of bits to identify highly correlated bit pairs.
We numerically compute the optimal partitioning for different realistic conditions.
Our results show that by employing the advantage distillation the transmission distance is significantly increased and thus can potentially improve the secret key rate of INI-QKD.
This in particular is most prominent in the presence of high polarization misalignment error rates and considerable phase mismatch, all without altering the experimental setup of the protocol.
\end{abstract}

\maketitle

%
%
\section{Introduction}%
Quantum Key Distribution (QKD) is a promising method that aims to establish information-theoretically secure communication between two distant parties, commonly referred to as Alice and Bob, even in the presence of an eavesdropper (Eve), by exploiting the laws of quantum mechanics~\cite{bennett2014quantum, ekert1991quantum}. With a securely shared key, Alice and Bob can subsequently exchange encrypted messages over a public channel without risking leakage to Eve.

Since the successful QKD experiment over a 32-cm free-space channel in 1992~\cite{bennett1992experimental}, considerable efforts have been devoted to enhancing the security and performance of QKD systems.  Remarkable advancements have been achieved, particularly in extending transmission distances and increasing key rates. A significant breakthrough in 2023 demonstrated secure QKD over an unprecedented distance of 1002 kilometers, highlighting the rapid progress in this field~\cite{liu2023experimental}. 

It is important to note, however, that practical implementations of QKD often differ from idealized theoretical models, leading to security vulnerabilities ~\cite{diamanti2016practical, sun2022review}. These vulnerabilities primarily stem from imperfections in physical components, such as modulators and detectors, which an eavesdropper can exploit through various side-channel attacks~\cite{xu2020secure}.
To address these issues, measurement-device-independent QKD (MDI-QKD) was introduced as a solution to eliminate the need for trusted detectors~\cite{lo2012measurement, braunstein2012side}. In this scheme, Alice and Bob send quantum signals to a third party, often called Charlie, who performs the measurements. The strength of MDI-QKD lies in its design, which requires no assumptions regarding the trustworthiness of the measurement devices. In other words, even if Charlie is under Eve's control, the protocol's security remains intact.

While MDI-QKD addresses detector vulnerabilities, it still encounters a fundamental limitation: the Pirandola-Laurenza-Ottaviani-Banchi (PLOB) bound~\cite{pirandola2017fundamental}. The PLOB bound defines the theoretical maximum key rate for point-to-point repeaterless QKD systems, based on the quantum channel's transmittance 
$\eta$, expressed as $\log_2(1-\eta)$. It illustrates the trade-off between transmission distance and secret key rate due to channel losses in such systems~\cite{takeoka2014fundamental, pirandola2017fundamental}. MDI-QKD remains constrained by this bound because it relies on two-photon interference, even though it is not a point-to-point protocol.

The introduction of twin-field QKD (TF-QKD) in 2018~\cite{lucamarini2018overcoming} represented a significant advancement in QKD by demonstrating the ability to surpass the PLOB bound using single-photon interference, thus eliminating the necessity for quantum repeaters~\cite{abruzzo2013quantum, amirloo2010quantum, mardani2020continuous}. This approach is based on similar principles MDI-QKD, where Alice and Bob send quantum signals to Charlie who performs the measurement. The results reveal the parity of the encoded bits without disclosing their actual values. Following the original TF-QKD protocol, several variants have been proposed to enhance security and performance further~\cite{Ma2018, Tamaki2018, Wang2018, Cui2019, Curty2019, Lin2018, Yin2019, Wang2020, Zeng2020, Xu2020, Curras2021, Zhang2020, Jiang2020, Zhang2021, Xu2021, Hu2022}.

A major challenge for high-speed QKD is achieving sufficiently high secret key rates, particularly over long distances. High-dimensional QKD (HD-QKD) addresses this by encoding information in qudits, $d$-dimensional quantum states where $d{>}2$, rather than qubits ($d{=}2$). Qudits can carry more information and are more resistant to noise and eavesdropping attacks compared to qubits~\cite{cerf2002security, bouchard2017high, pirandola2020advances}. Although the information density per mode decreases logarithmically with $d$, HD-QKD provides significant advantages in key rate and noise tolerance\cite{pirandola2020advances}. An example of this approach is the interfering-or-not-interfering QKD (INI-QKD)~\cite{yu2024interfering}, which encodes two bits of information per quantum state using the polarization and phase of coherent states. INI-QKD has been shown to outperform all variants of TF-QKD in terms of transmission distance and secret key rates under negligible phase mismatch ($\delta$) and polarization misalignment error rates ($e_d$), positioning it as a promising candidate for future QKD systems.

Nonetheless, enhancing the transmission distance without quantum repeaters and increasing the error rate tolerance should be resolved in practical QKD schemes.
In addition to introducing novel protocols, developing post-processing methods are essential to improve the performance of a QKD protocol without significant costs.
The conventional post-processing approaches in QKD typically include error correction and privacy amplification~\cite{bennett1988privacy, brassard1993secret}. Advanced methods based on entanglement distillation extend these approaches~\cite{He2022, Kumar2023, mardani2020continuous, Rozpedek2018}.
Among the others are the advantage distillation (AD) methods, initially proposed in classical cryptography theory~\cite{Maurer},
which includes methods based on repetition codes to improve the tolerable error rate~\cite{renner2008security, bae2007key, tan2020advantage, kraus2007security}.
Two-way classical communications, another variant of the AD method, have been used to enhance the key rate and the error tolerance~\cite{Gottesman, Ma}.
The AD method as a pre-processing of raw keys, before standard information reconciliation and privacy amplification, has been previously utilized in various QKD protocols, see e.g. Refs.~\cite{murta2020key, li2022improving, wang2022phase, jiang2023improving, zhou2024sending, zhang2024discrete, liu2023mode, zhu2023reference, hu2023practical, wang2024improving, luo2024practical, Li2023improving}.

This paper proposes enhancing the maximal transmission distance and the error rate tolerance of the INI-QKD protocol by applying the AD method, defined as a classical operation before the post-processing phase that uses two-way communication instead of the traditional one-way error correction approach~\cite{renner2008security}.
Our numerical results indicate that an AD-applied INI-QKD can effectively tolerate high levels of system misalignment errors, resulting in noticeable improvements in both the secret key rate and transmission distance. By segmenting the raw key bits into smaller blocks of size $b\leq3$, the INI-QKD protocol can surpass the PLOB bound even with polarization misalignment as high as \(e_d = 0.50\) or phase mismatch as large as \(\delta = 0.25\). 

The rest of this paper is organized as follows: Sec.~\ref{Security of QKD with AD} offers an overview of the security of AD-applied QKD. In Sec.~\ref{AD-applied INI-QKD}, we introduce the INI-QKD protocol enhanced with the AD method. The numerical results are presented and discussed in Sec.~\ref{simulations}. Finally, Sec.~\ref{conc} provides the concluding remarks.

\section{Security of AD-applied QKD}\label{Security of QKD with AD}
In a standard QKD protocol, Alice's and Bob's raw keys may not be perfectly correlated due to noise, eavesdropping, or imperfections in the quantum channel.
They thus need to agree on a shared key by using public communication, which in principle can be done bit by bit.
Nevertheless, to enhance the efficiency the core step of the AD method involves partitioning the raw key into small blocks to distinguish highly correlated bit pairs from less correlated information.
Therefore, the correlation between the raw keys will be increased.
It is crucial to note that the AD method does not alter the hardware devices involved in quantum state preparation and measurement, the quantum step of a practical QKD protocol.
Indeed it serves as a pre-processing step that exclusively modifies the classical post-processing phase to enhance the performance, which can be conveniently applied to different practical QKD protocols.
In this section, we prove the security of an AD-applied QKD protocol.

In an entanglement-based QKD scheme, Alice prepares the Bell state \(\frac{1}{\sqrt{2}}(\lvert 00 \rangle + \lvert 11 \rangle)\) and sends the second qubit to Bob through the noisy quantum channel, while storing the first one.
Alice and Bob then perform measurements in either the \(Z\) or \(X\) basis, where the \(Z\) basis consists of \(\lvert 0 \rangle\) and \(\lvert 1 \rangle\) states, and the \(X\) basis consists of \(\lvert + \rangle = \frac{1}{\sqrt{2}}(\lvert 0 \rangle + \lvert 1 \rangle)\) and \(\lvert - \rangle = \frac{1}{\sqrt{2}}(\lvert 0 \rangle - \lvert 1 \rangle)\) states.
The final state shared between Alice and Bob can be expressed in the Bell states basis as
\begin{equation}
\begin{aligned}
\sigma_{AB} = \lambda_0 \ket{\Phi_0} \langle \Phi_0 \rvert + \lambda_1 \lvert \Phi_1 \rangle \langle \Phi_1 \rvert \\
+ \lambda_2\lvert \Phi_2 \rangle \langle \Phi_2 \rvert + \lambda_3 \lvert \Phi_3 \rangle \langle \Phi_3 \rvert,
\end{aligned}
\end{equation}
where the Bell states are defined as 
\begin{equation}
\begin{aligned}
|\Phi_0\rangle &= \frac{1}{\sqrt{2}}(|00\rangle + |11\rangle), & |\Phi_1\rangle &= \frac{1}{\sqrt{2}}(|00\rangle - |11\rangle), \\
|\Phi_2\rangle &= \frac{1}{\sqrt{2}}(|01\rangle + |10\rangle), & |\Phi_3\rangle &= \frac{1}{\sqrt{2}}(|01\rangle - |10\rangle).
\end{aligned}
\end{equation}
The presence of noise within the quantum channel and potential eavesdropping by Eve, introduces errors in both the $\mathbf{Z}$ and $\mathbf{X}$ bases, usually referred to as quantum bit and phase errors, respectively.
The corresponding error rates, denoted as \( E^{\text{bit}} \) and \( E^{\text{ph}} \), are subject to the following constraints~\cite{renner2008security}:
\begin{equation}
    \lambda_2 + \lambda_3 = E^\text{bit}, \quad \lambda_1 + \lambda_3 = E^\text{ph}.
\end{equation}
For a trace-preserving quantum channel, commonly employed as a model for quantum channels in QKD, the parameters \(\lambda_0\), \(\lambda_1\), \(\lambda_2\), and \(\lambda_3\) must satisfy the condition
\begin{equation}
    \lambda_0 + \lambda_1 + \lambda_2 + \lambda_3 = 1.
\end{equation}
The secret key rate of a QKD protocol is given by~\cite{renner2008security}
\begin{align}
R &\geq \min_{\lambda_0, \lambda_1, \lambda_2, \lambda_3} \left[ S(A|E) - H(A|B) \right] \nonumber\\
  &= \min_{\lambda_0, \lambda_1, \lambda_2, \lambda_3} \Bigg[ 1 - (\lambda_0 + \lambda_1)H\left(\frac{\lambda_0}{\lambda_0 + \lambda_1}\right) \\
  &\quad - (\lambda_2 + \lambda_3)H\left(\frac{\lambda_2}{\lambda_2 + \lambda_3}\right) - H(\lambda_0 + \lambda_1) \Bigg], \nonumber
\end{align}
where \(E\) represents Eve's ancillary system, while \(A\) and \(B\) denote Alice's and Bob's systems, respectively.
Here, $S(\rho) = -\text{tr}(\rho \log \rho)$ is the von Neumann entropy and the conditional entropy is given by $S(A|E) = S(A, E) - S(E)$.
Furthermore, $H(x)= -x \log(x) - (1 - x) \log(x)$ is the Shannon binary entropy.
Note that we assume Eve can freely choose the optimal \(\lambda_i\) to maximize her eavesdropping, as long as \(\lambda_i\) is constrained by the error rates.

To apply the AD method, Alice and Bob divide their sifted keys into blocks of $b$ bits \(\{x_1, x_2, \ldots, x_b\}\) and \(\{y_1, y_2, \ldots, y_b\}\), respectively.
Depending on a randomly chosen bit \(c \in \{0, 1\}\), Alice transmits the message \(m = \{m_1, m_2, \ldots, m_b\} = \{x_1 \oplus c, x_2 \oplus c, \ldots, x_b \oplus c\}\) to Bob over an authenticated classical channel.
The block is accepted if and only if Bob announces that the result of \(\{m_1 \oplus y_1, m_2 \oplus y_2, \ldots, m_b \oplus y_b\}\) equals \(\{0, 0, \ldots, 0\}\) or \(\{1, 1, \ldots, 1\}\).
If accepted, they retain the first bit of their blocks, $x_1$ and $y_1$, as their processed key.
As Alice and Bob can manipulate the parameter $b$ in the AD method, they have the ability to select the optimal value of $b$ to enhance the secure key rate.

The success probability of the AD method for a block of size \(b\) is then given by
\begin{equation}
    p_{\text{succ}} = (1-E^\text{bit})^{b} + (E^\text{bit})^{b} = (\lambda_0 + \lambda_1)^b + (\lambda_2 + \lambda_3)^b.
    \label{p_succ}
\end{equation}
Therefore, the resulting quantum state shared between Alice and Bob can be represented as
\begin{align*}
        \tilde{\sigma}_{AB} &= \tilde{\lambda}_0 \lvert \Phi_0 \rangle \langle \Phi_0 \rvert + \tilde{\lambda}_1 \lvert \Phi_1 \rangle \langle \Phi_1 \rvert \\
        &+ \tilde{\lambda}_2 \lvert \Phi_2 \rangle \langle \Phi_2 \rvert + \tilde{\lambda}_3 \lvert \Phi_3 \rangle \langle \Phi_3 \rvert,
\end{align*}
where \(\tilde{\lambda}_0\), \(\tilde{\lambda}_1\), \(\tilde{\lambda}_2\), and \(\tilde{\lambda}_3\) are given in the following
\begin{align}
\tilde{\lambda}_0 &= \frac{(\lambda_0 + \lambda_1)^b + (\lambda_0 - \lambda_1)^b}{2p_{\text{succ}}}, & \tilde{\lambda}_1 &= \frac{(\lambda_0 + \lambda_1)^b - (\lambda_0 - \lambda_1)^b}{2p_{\text{succ}}}, \nonumber\\
\tilde{\lambda}_2 &= \frac{(\lambda_2 + \lambda_3)^b + (\lambda_2 - \lambda_3)^b}{2p_{\text{succ}}}, & \tilde{\lambda}_3 &= \frac{(\lambda_2 + \lambda_3)^b - (\lambda_2 - \lambda_3)^b}{2p_{\text{succ}}}.
\label{tildelambdas}
\end{align}
After applying the AD method to all $b$-bit blocks of Alice and Bob's data, they proceed with error correction and privacy amplification to obtain the final secret key. The rate is then expressed as:
\begin{equation}
\begin{split}
\tilde{R} &\geq \max_{b} \min_{\lambda_0, \lambda_1, \lambda_2, \lambda_3} 
\frac{1}{b} p_{\text{succ}} \Big[
1 - (\tilde{\lambda}_0 + \tilde{\lambda}_1) H\big(\frac{\tilde{\lambda}_0}{\tilde{\lambda}_0 + \tilde{\lambda}_1}\big) \\
&\quad - (\tilde{\lambda}_2 + \tilde{\lambda}_3) H\big(\frac{\tilde{\lambda}_2}{\tilde{\lambda}_2 + \tilde{\lambda}_3}\big)
- H(\tilde{\lambda}_0 + \tilde{\lambda}_1) \Big].
\end{split}
\label{adkeyrate}
\end{equation}
\section{AD-applied INI-QKD}\label{AD-applied INI-QKD}
\begin{figure}
    \centering
    \includegraphics[scale=0.29]{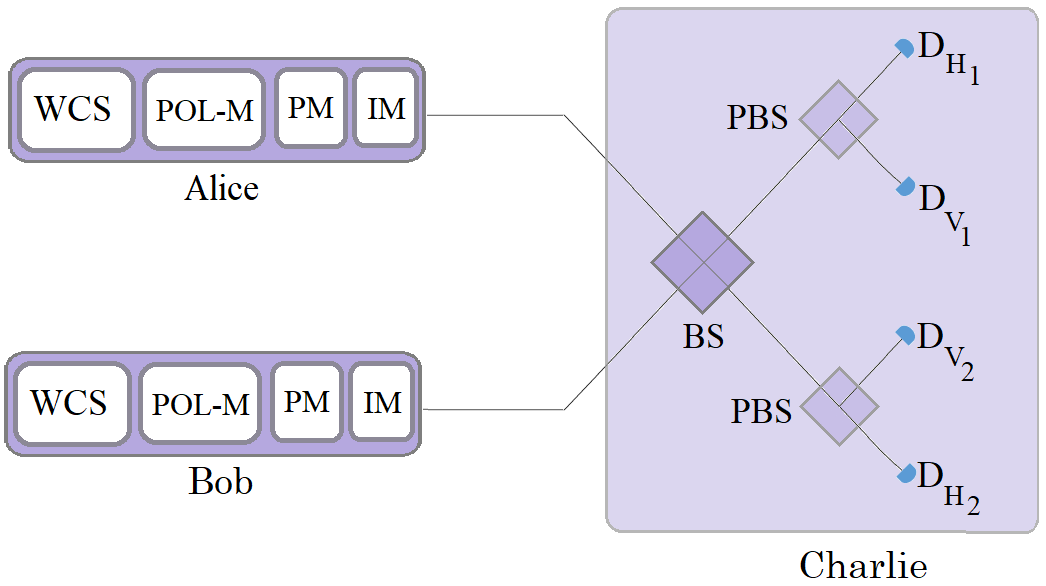}
    \caption{
    Schematic setup of the INI-QKD: WCS, weak coherent pulse source; POL-M, polarization modulator; PM, phase modulator; IM, intensity modulator; BS, beam splitter; PBS, polarization beam splitter.
    }
    \label{fig:setupfig}
\end{figure}

The general scheme of the AD-applied INI-QKD protocol is shown in Fig.~\ref{fig:setupfig}, and the procedure runs as follows:

\textbf{(1) State Preparation.} In each round of the protocol, Alice and Bob independently choose to prepare a polarized coherent state using either the $\mathbf{X}$ (diagonal) basis or the $\mathbf{Z}$ (rectilinear) basis. They select a polarization bit, $\kappa_{a(b)}^{\text{pol}}$, and a phase bit, $\kappa_{a(b)}^{\text{ph}}$, each one randomly and with equal probability from $\{0, 1\}$.

For the $\mathbf{X}$ basis, if $\kappa_{a(b)}^{\text{pol}} = 0$, Alice (Bob) prepares the state $\ket{\sqrt{\mu} e^{i \pi \kappa_{a(b)}^{\text{ph}}}}_{A_{+}(B_{+})}$. Conversely, if $\kappa_{a(b)}^{\text{pol}} = 1$, the state becomes $\ket{\sqrt{\mu} e^{i \pi \kappa_{a(b)}^{\text{ph}}}}_{A_{-}(B_{-})}$. Again, this process results in four possible states, each defined by the chosen polarization and phase bits.

Similarly, for the $\mathbf{Z}$ basis, Alice (Bob) prepares the state $\ket{\sqrt{\mu_{A(B)}} e^{i \pi\kappa_{a(b)}^{\text{ph}}}}_{A(B)_H}$ when $\kappa_{a(b)}^{\text{pol}} = 0$, or $\ket{\sqrt{\mu_{A(B)}} e^{i \pi \kappa_{a(b)}^{\text{ph}}}}_{A(B)_V}$ when $\kappa_{a(b)}^{\text{pol}} = 1$. This leads to a set of four states associated with different combinations of polarization and phase choices.

Note that the intensities $\mu_{A(B)}$ in the $\mathbf{Z}$ polarization basis are randomly chosen from a predefined set rather than being fixed. Additionally, phase-locking technique is crucial to ensure that Alice's and Bob's quantum states share a common global phase, enabling meaningful interference at Charlie's site.

\textbf{(2) Measurement.} 
Alice and Bob send their prepared quantum states to an Charlie, whose role is to correlate the states through interference detection. He records the measurement results for each round and announces them. Only specific measurement results, listed below, are retained, while all others are discarded.
\begin{enumerate}
    \item The $X_1$ event: only one of detectors $D_{H_1}$ or $D_{H_2}$ clicks.
    \item The $X_2$ event: detectors ($D_{H_1}$, $D_{V_1}$) or ($D_{H_2}$, $D_{V_2}$) click simultaneously. 
    \item The $X_3$ event: detectors ($D_{H_1}$, $D_{V_2}$) or ($D_{H_2}$, $D_{V_1}$) click simultaneously. 
\end{enumerate}

\textbf{(3) Announcement.} The protocol is repeated multiple times to collect sufficient data for key generation and error analysis. Afterwards, Alice and Bob announce their basis choices for each round through a public classical channel. They then select only the instances where both chose the same polarization basis. The measurements taken in the $\mathbf{X}$ basis are used to extract the raw key bits, while the measurements in the $\mathbf{Z}$ basis serve to detect any potential eavesdropping by Eve.

\textbf{(4) Parameter Estimation.} A portion of the raw key bits is sampled to estimate the overall gain and the bit error rate. If the bit error rate is below a predetermined threshold, they proceed. Otherwise, they abort the protocol.

\textbf{(5) AD.} Alice and Bob apply the AD method to their blocks of $b$ bits in their raw keys, thereby generating highly-correlated processed key bits.

\textbf{(6) Post-processing.} Alice and Bob perform error correction and privacy amplification to generate the final secret key. 
\section{The secret key rate of AD-applied INI-QKD}\label{SKR of AD-applied INI-QKD}
In the INI-QKD protocol, the overall secret key rate $R$ is given by the sum of the secret key rates of the three effective events recorded in the second step of the protocol:
\begin{equation}
R = \sum_{i=1}^{3} R^{X_i} = \sum_{i=1}^{3} Q^{X_i} \left[ 1 - H\left(E_{\text{ph}}^{X_i}\right) - I_E^{U} - f H\left(E_{\text{bit}}^{X_i}\right)\right],
\label{R}
\end{equation}
where \( Q^{X_i} \) is the gain, while \( E_{\text{ph}}^{X_i} \) and \( E_{\text{bit}}^{X_i} \) denote the phase and quantum bit error rates of the effective event \( X_i \), respectively.
Here, \( f \) is the error correction efficiency. Additionally, \( I_E^U \) shows the upper bound on the mutual information that Eve can obtain via a beam-splitting attack which is derived from the maximum probability of Eve unambiguously discriminating the prepared states and is found as
\cite{yu2024interfering}
\begin{equation}
I_E^U = 1 - \frac{1}{9} \left[ e^{-2(1-\eta)\mu} + 2e^{-(1-\eta)\mu} \right]^2.
\end{equation}
Further details are provided in Appendix A.

\begin{figure*}
        \includegraphics[width=0.9\columnwidth]{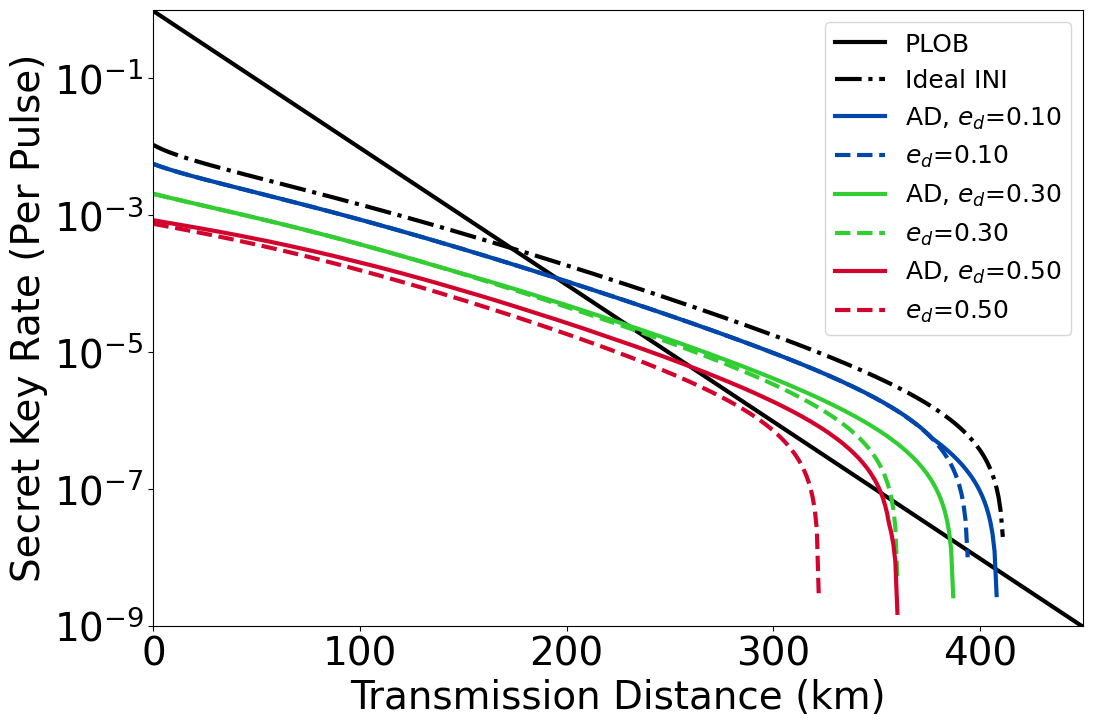}
        \put(-185,25){(a)}
        \includegraphics[width=0.85\columnwidth]{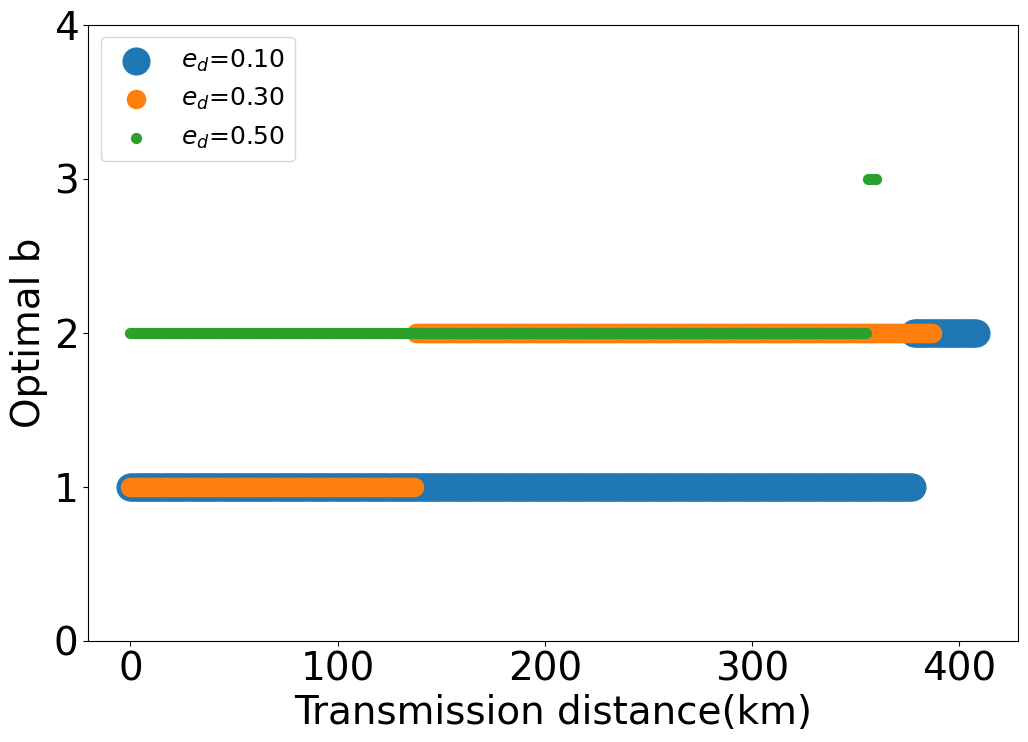}
        \put(-185,25){(b)}
        \caption{(a) Secret key rates with respect to the distance for the AD-applied INI-QKD protocol (colored solid lines) and the original INI-QKD protocol (dashed lines) for $e_d= 0.10$, $0.30$ and $0.50$. The black solid line represents the PLOB bound, and the dot-dashed curve represents the ideal INI-QKD protocol where $e_d=0$.
        (b) The corresponding optimal $b$ values with respect to the transmission distance in AD-applied INI-QKD.}
        \label{fig:eds}
\end{figure*}

When applying the AD to all the effective events of the INI-QKD protocol, the parameters are modified as
\begin{equation}
    E_{\text{bit}}^{X_i} = \lambda_2^{X_i} + \lambda_3^{X_i}, \quad E_{\text{ph}}^{X_i} = \lambda_1^{X_i} + \lambda_3^{X_i}.
\end{equation}
From Eq.~(\ref{p_succ}), the success probability of the effective event $X_i$ is given by
\begin{equation}
    p^{X_i}_{\text{succ}} = \left( E_{\text{bit}}^{X_i} \right)^{b_i} + \left( 1 - E_{\text{bit}}^{X_i} \right)^{b_i}.
\end{equation}
Here, \( b_i \) represents the optimal block size when applying the AD method to the bits extracted from the effective event \( X_i \).
As a result of applying the AD method, the quantum bit error rate of the effective event $X_i$ will also be modified to
$\tilde{E}_{\text{bit}}^{X_i} = \left( E_{\text{bit}}^{X_i} \right)^{b_i}/p^{X_i}_{\text{succ}}$.
One then finds the Bell state probabilities corresponding to each event from Eqs.~\eqref{tildelambdas}
\begin{equation} 
\begin{aligned}
\tilde{\lambda}^{X_i}_{0} &= \frac{(\lambda^{X_i}_0 + \lambda^{X_i}_1)^{b_i} + (\lambda^{X_i}_0 - \lambda^{X_i}_1)^{b_i}}{2p^{X_i}_\text{succ}}, \\
\tilde{\lambda}^{X_i}_{1} &= \frac{(\lambda^{X_i}_0 + \lambda^{X_i}_1)^{b_i} - (\lambda^{X_i}_0 - \lambda^{X_i}_1)^{b_i}}{2p^{X_i}_\text{succ}}, \\
\tilde{\lambda}^{X_i}_{2} &= \frac{(\lambda^{X_i}_2 + \lambda^{X_i}_3)^{b_i} + (\lambda^{X_i}_2 - \lambda^{X_i}_3)^{b_i}}{2p^{X_i}_\text{succ}}, \\
\tilde{\lambda}^{X_i}_{3} &= \frac{(\lambda^{X_i}_2 + \lambda^{X_i}_3)^{b_i} - (\lambda^{X_i}_2 - \lambda^{X_i}_3)^{b_i}}{2p^{X_i}_\text{succ}}.
\end{aligned}
\end{equation}
By plugging back these in Eqs.~\eqref{adkeyrate} and \eqref{R} the secret key rate of AD-applied INI-QKD is then found as
\begin{align}
\tilde{R} =& \sum_{i=1}^{3} \max_{b_i} \Bigg\{\min_{\{\lambda^{X_i}\}} \bigg\{\frac{1}{b_i} Q^{X_i}_{\mu} p_\text{succ}^{X_i} \bigg[ 1 - (\tilde{\lambda}^{X_i}_1 + \tilde{\lambda}^{X_i}_2) H \Big( \frac{\tilde{\lambda}^{X_i}_2}{\tilde{\lambda}^{X_i}_2 + \tilde{\lambda}^{X_i}_3} \Big) \nonumber\\
&- (\tilde{\lambda}^{X_i}_2 + \tilde{\lambda}^{X_i}_3) H \Big( \frac{\tilde{\lambda}^{X_i}_2}{\tilde{\lambda}^{X_i}_2 + \tilde{\lambda}^{X_i}_3} \Big)  - f H\left(\tilde{E}_\text{bit}^{X_i}\right)
 - I_{E}^{U} \bigg]\bigg\}\Bigg\}.
\end{align}
In the next section, we numerically compute the secret key rate from this equation for different conditions.

\section{Numerical Results}\label{simulations}
To evaluate the performance of AD-applied INI-QKD, we assume the following parameters: the detection efficiency \( \eta_d \) of the single-photon detectors is \( 14.5\% \), the dark count rate \( p_d \) is \( 8 \times 10^{-8} \), the efficiency of error correction  \( f \) is \( 1.15 \), and the channel loss \( \alpha \) is \( 0.2 \, \text{dB/km} \).
The size \( b \) of each split block in AD is optimized. 

We assess the AD-applied INI-QKD protocol under three distinct scenarios, comparing it to the original INI-QKD protocol.
First, we examine the case where the phase mismatch ($\delta$) is negligible, but the polarization misalignment error rate ($e_d$) is significant.
Second, we consider the opposite scenario, where $e_d$ is negligible, but \( \delta \) is substantial.
Finally, we analyze situations where both \( \delta \) and \( e_d \) are considerable and can affect the protocol.
In our following computations for the sake of simplicity we homogeneously optimize the block sizes for all events.
That is, we maximize the secret key rate by assuming $b_i=b$.
\begin{figure*}
    \centering
    \includegraphics[width=0.9\columnwidth]{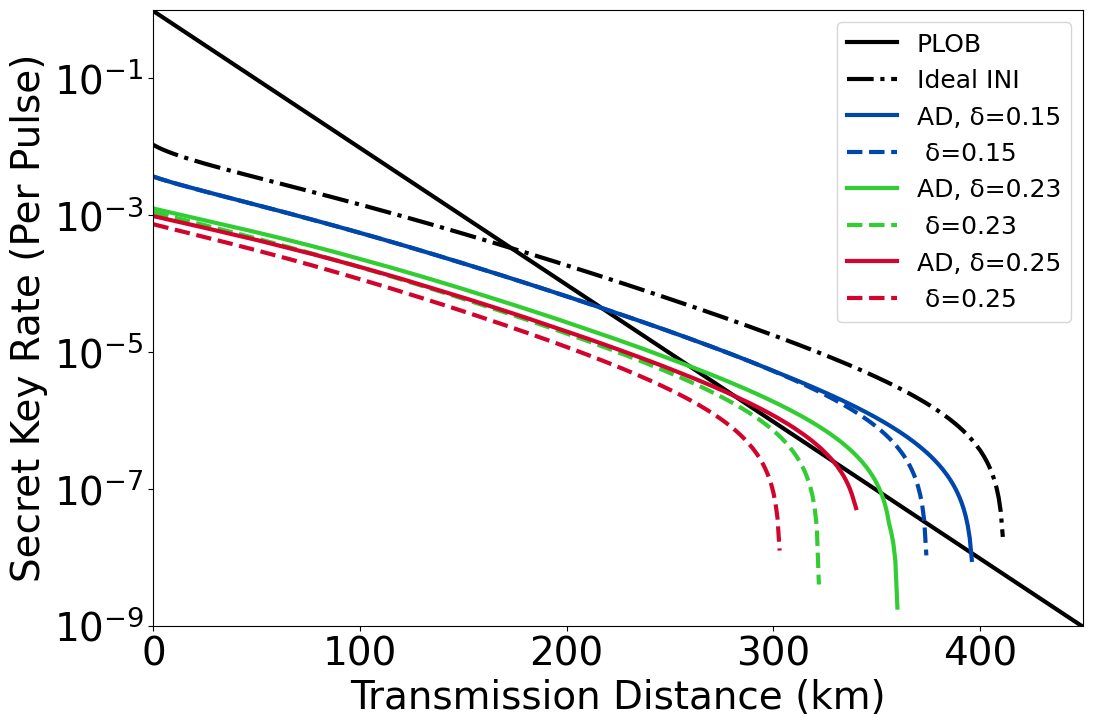}
    \put(-185,25){(a)}
    \includegraphics[width=0.85\columnwidth]{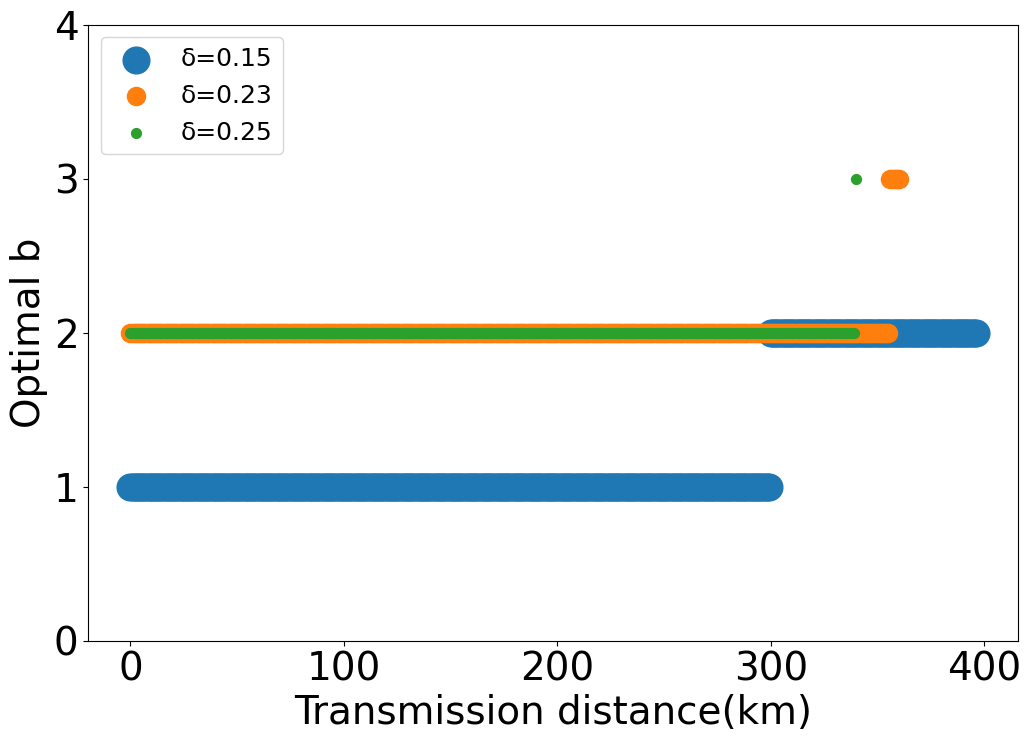}
    \put(-185,25){(b)}
    \caption{(a) The secret key rate with respect to the transmission distance for the AD-applied (colored solid lines) and original (dashed lines) INI-QKD protocols with phase mismatch values of $\delta = 0.15, 0.23,$ and $0.25$. The black solid line represents the PLOB bound, and the dot-dashed curve shows the ideal INI-QKD protocol where $\delta=0$.(b) The corresponding optimal $b$ with respect to the transmission distance in AD-applied INI-QKD.}
    \label{fig:deltas}
\end{figure*}

\subsection{Negligible phase mismatch}\label{subsecA}
First, we examine the effect of polarization misalignment error on the protocol and the improvements brought by our AD method to the INI-QKD. Specifically, we consider three scenarios: \(e_d = 0.10\), \(0.30\), and \(0.50\), all with \(\delta = 0\). 

In Fig.~\ref{fig:eds}(a), the secret key rate is plotted against the transmission distance for each case. The simulation results demonstrate that the AD method consistently enhances the maximum achievable transmission distance between Alice and Bob for all considered values of the polarization misalignment error rate \(e_d\). This improvement becomes more pronounced as \(e_d\) increases, leading to both an increase in the secret key rate and an extension of the transmission distance. 

For example, when \(e_d = 0.50\), the original INI-QKD protocol fails to surpass the PLOB bound. Nevertheless, by incorporating the AD method into the protocol, the key rate successfully exceeds this bound, demonstrating the effectiveness of the AD method in handling scenarios with high polarization misalignment errors. In this case, the maximum transmission distance increases from 323 km (without AD) to 361 km (with AD). Similarly, for \(e_d = 0.30\), the maximum transmission distance extends from 361 km to 388 km, and for \(e_d = 0.10\), it increases from 395 km to 409 km when AD is applied. Although the relative improvement is more significant for higher error rates, even at lower error rates like \(e_d = 0.10\), the AD method provides noticeable benefits. As the system approaches ideal conditions, the improvements provided by AD become less significant, suggesting declining advantages in such low-error scenarios.

Fig.~\ref{fig:eds}(b) presents the optimal block size \(b\)with respect to the transmission distance for each polarization misalignment error rate. For smaller error rates such as \(e_d = 0.10\) and \(e_d = 0.30\), the optimal \(b\) remains at \(1\) over a large portion of the transmission distance. This suggests that in these cases, the application of the AD method is not immediately required to enhance performance, as the errors remain within a tolerable range. 

Conversely, for high polarization misalignment error rates such as \(e_d = 0.50\), the optimal \(b\) initially exceeds \(1\), suggesting that applying AD right from the start can enhance the key rate, even at short transmission distances. This result highlights that, in scenarios with higher polarization misalignment, immediate block splitting not only improves the key rate but also extends the maximum achievable transmission distance. Employing AD early on allows the system to better handle the effects of increased errors, ensuring more robustness over longer distances.

Furthermore, as the transmission distance increases, the optimal \(b\) rises for all error rates. This indicates the need for larger blocks to manage accumulating errors and emphasizing the importance of dynamic adaptation in key rate optimization.

\subsection{Negligible polarization misalignment error rate}\label{subsecB}
Next, we compare the AD-applied INI-QKD protocol with the original INI-QKD protocol when the errors arise solely from phase mismatch. For this comparison, we set \( e_d = 0 \) and consider phase mismatch errors of \( \delta = 0.15 \), \( 0.23 \), and \( 0.25 \). 

Fig.~\ref{fig:deltas}(a) illustrates the secret key rate as a function of transmission distance under these conditions. The results clearly show that the AD-applied protocol enhances the key rate across all values of \( \delta \). Notably, at \( \delta = 0.25 \), the original protocol fails to surpass the PLOB bound, limiting its secure communication range. In contrast, the AD-applied protocol successfully exceeds this bound, enabling secure communication over longer distances. For instance, the maximum transmission distance increases from 304 km (without AD) to 341 km (with AD). Similarly, for \( \delta = 0.23 \), the distance extends from 323 km to 361 km, and for \( \delta = 0.15 \), it increases from 375 km to 397 km with the use of AD. Even at lower phase mismatch values like \( \delta = 0.15 \), where the original protocol performs relatively well, the AD method still offers a noticeable improvement in the transmission distance.

Fig.~\ref{fig:deltas}(b) shows the corresponding optimal block size \( b \) with respect to the transmission distance. For \( \delta = 0.15 \), the optimal \( b \) remains at 1 over a large portion of the distance, indicating that block splitting is not immediately necessary. However, as the distance grows, the optimal \( b \) increases to 2, suggesting that identifying highly correlated raw bits through AD becomes beneficial. For higher phase mismatch values, such as \( \delta = 0.23 \) and \( \delta = 0.25 \), the optimal \( b \) starts off greater than 1 even at short distances, indicating that an AD-applied protocol is required from the start to better manage the increased phase mismatch and enhance the secret key rate.
\begin{figure*}
    \centering
    \includegraphics[width=0.9\columnwidth]{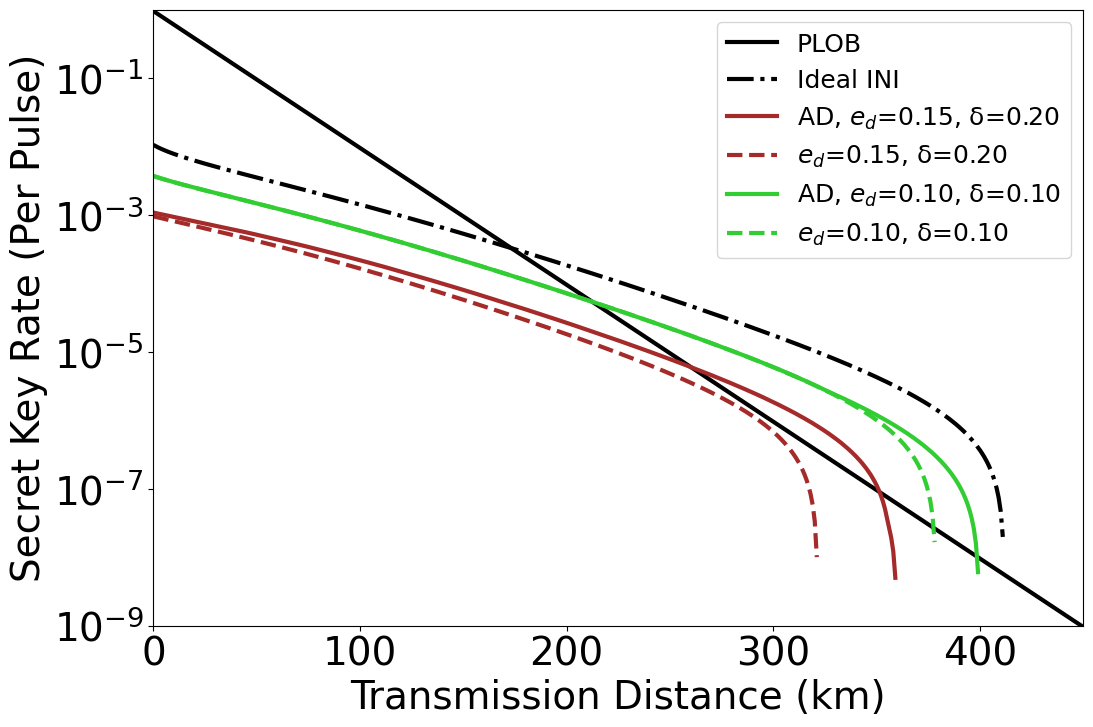}
    \put(-185,25){(a)}
    \includegraphics[width=0.85\columnwidth]{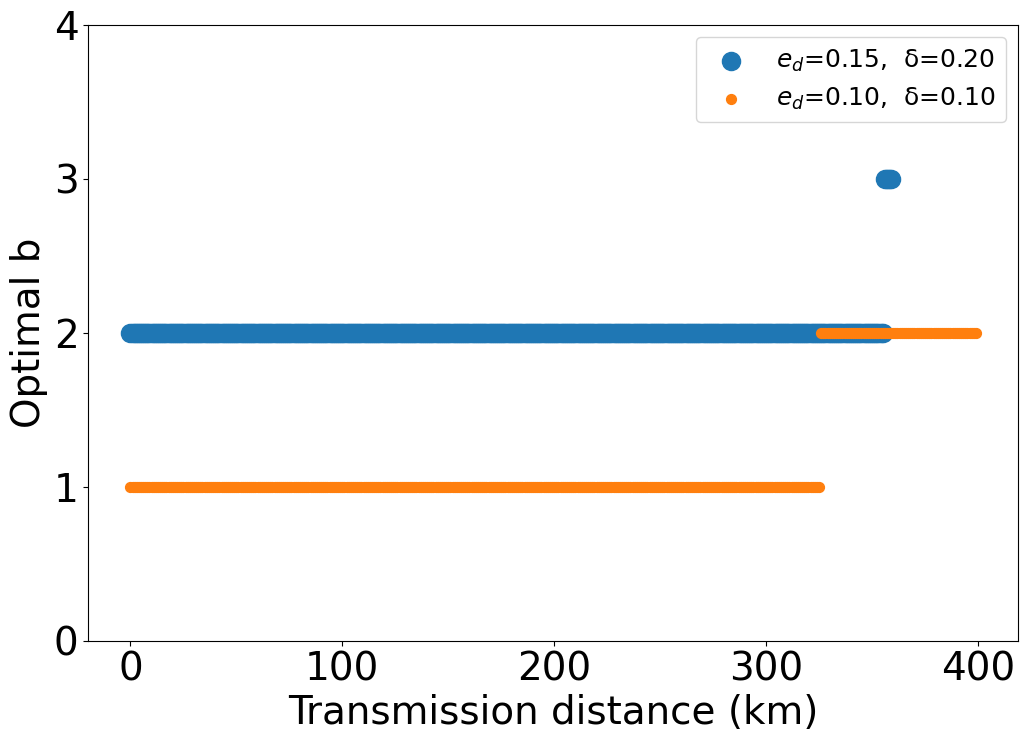}
    \put(-185,25){(b)}
    \caption{(a) The secret key rate with respect to transmission distance between Alice and Bob for $e_d = 0.15$ and $\delta = 0.20$, as well as $e_d = 0.10$ and $\delta = 0.10$. The black line represents the PLOB bound, with dashed curves depicting the performance of the original INI-QKD protocol and solid curves showing the results of the AD-applied INI-QKD protocol. (b) The optimal $b$ as a function of transmission distance for $e_d = 0.15$ and $\delta = 0.20$, as well as for $e_d = 0.10$ and $\delta = 0.10$.}
    \label{fig:eanddeltas}
\end{figure*}

\subsection{AD-applied INI-QKD with phase mismatch and polarization misalignment errors}\label{subsecC}
Finally, we examine the scenario where both phase mismatch and polarization misalignment errors are considerable. Fig.~\ref{fig:eanddeltas}(a) shows the secret key rate with respect to the transmission distance for two cases: \( \delta = 0.20 \), \( e_d = 0.15 \) and \( \delta = 0.10 \), \( e_d = 0.10 \).

The results indicate that applying the AD method provides noticeable improvements in both the secret key rate and transmission distance. For instance, when \( \delta = 0.20 \) and \( e_d = 0.15 \), the original INI-QKD protocol cannot exceed the PLOB bound, limiting the maximum transmission distance to 322 km. In contrast, the AD-applied protocol not only surpasses this bound but also extends the maximum transmission distance to 360 km. Similarly, for the lower error rates of \( \delta = 0.10 \) and \( e_d = 0.10 \), the original protocol achieves a maximum distance of 379 km, while the AD-applied protocol increases this range to 400 km. This consistent performance boost highlights the robustness of the AD method in managing combined phase mismatch and polarization misalignment errors, even under challenging conditions.

Fig.~\ref{fig:eanddeltas}(b) depicts the corresponding optimal clustering factor \( b \) with respect to the transmission distance. When both errors are present, the optimal \( b \) remains primarily at 2 across a wide range of distances for most of the scenarios. This behavior suggests that initiating the protocol with a block size of 2 is crucial to counteract the combined effects of phase mismatch and polarization misalignment and achieve a higher secret key from the outset.

Interestingly, for the lower error scenario (\( \delta = 0.10 \), \( e_d = 0.10 \)), the optimal \( b \) stays at 1 for most of the transmission range, only increasing slightly towards the longer distances. This indicates that in relatively less noisy environments, the system can maintain a stable key rate without immediately resorting to larger block sizes. However, as the transmission distance grows, the slight increase in \( b \) signals the necessity of block splitting to handle accumulating errors over longer distances.
\section{Conclusion}\label{conc} 
In conclusion, we proposed the advantage distillation (AD)-applied INI-QKD protocol and analyzed its performance considering practical imperfections like polarization misalignment and phase mismatch. Our results show that the AD method improves both the secret key rate and transmission distance, even under high error rates (up to \(e_d = 0.50\) and \(\delta = 0.25\)). For example, AD extends the maximum transmission distance from 323 km to 361 km at \(e_d = 0.50\) and from 304 km to 341 km at \(\delta = 0.25\). The optimal block size \(b\) adapts to varying conditions, starting at 1 for low errors and increasing to 2 or more as errors rise or distances grow. This adaptability allows the protocol to effectively handle errors, maintaining secure communication. Importantly, AD requires no hardware changes, making it a practical enhancement for existing QKD systems.

\appendix\label{Appendix A}
\section{Parameters of secret key rate}
In this appendix provide the details of Eq.~(\ref{R}) from the main text, accounting for practical imperfections such as polarization misalignment and phase mismatch. Polarization misalignment, assumed to result from the rotation of state polarizations during transmission, is modeled by the following unitary operator \cite{yu2024interfering, xu2013practical}
\begin{equation}
    U_{A(B)} = \begin{pmatrix} \cos \theta_{A(B)} & -\sin \theta_{A(B)} \\ \sin \theta_{A(B)} & \cos \theta_{A(B)} \end{pmatrix}
    \label{matrix}
\end{equation}
where $\theta_{A(B)}$ is the polarization-rotation angle. Due to the symmetry of the channels, we assume that \( \sin^2(\theta_A) = \sin^2(\theta_B) = e_d/2 \). Based on this assumption, there are two possible cases: symmetric \( \theta_A = \theta_B = \arcsin \sqrt{e_d / 2} \), or antisymmetric \( \theta_A = -\theta_B = \arcsin \sqrt{e_d / 2} \). For simplicity, we have only considered the symmetric situation in this paper. 

To account for the phase mismatch, we introduce phase shifts $\phi_A$ and $\phi_B$ for Alice's and Bob's states, respectively.
The phase mismatch is quantified by the dimensionless parameter $\delta = \phi_{\delta} / \pi$, where $\phi_{\delta} = \phi_B - \phi_A$.

The intensities received by each detector are denoted as \( I_{ijH(V)_{k}mn} \), where \( i \) and \( j \) represent the polarization states chosen by Alice and Bob, respectively. 
The subscript \( H(V)_{k} \) with $k=1,2$ refers to the specific detector receiving the intensity, while \( m \) and \( n \) indicate the respective phase bits selected by Alice and Bob.
Based on Eq.~(\ref{matrix}) and considering that the quantum channels between Alice-Charlie and Bob-Charlie behave as beam splitters with a transmittance \( \eta = 10^{-\alpha l/20} \), the intensities at each detector can be calculated for different polarization states and phase bits. 
Here, \( \alpha \) represents the channel loss, and \( l \) is the transmission distance between Alice and Bob.
The explicit form of the intensities are too cumbersome to be reported here.
Having the intensities determined, one can proceed to calculate the parameters for each effective event.
Note that in the following \( \Bar{a} \) and \( \Bar{b} \) denote the binary XOR of the phase bits \( a \) and \( b \), respectively.

\subsection{\texorpdfstring{$X_1$}{X\_1} event}
The gain of detector \( H_1 \) (or \( H_2 \)) for the phase bits \( a \) and \( b \) (or their complements \( \Bar{a} \) and \( \Bar{b} \)) chosen by Alice and Bob is denoted by \( Q^{H_1}_{ab} = Q^{H_1}_{\Bar{a}\Bar{b}} \) (or \( Q^{H_2}_{ab} = Q^{H_2}_{\Bar{a}\Bar{b}} \)). 

Similarly, the phase error rate when detector \( H_1 \) (or \( H_2 \)) clicks is given by \( E^{\text{ph},H_1}_{ab} = E^{\text{ph},H_1}_{\Bar{a}\Bar{b}} \) (or \( E^{\text{ph},H_2}_{ab} = E^{\text{ph},H_2}_{\Bar{a}\Bar{b}} \)). 

The bit error rate for the \( X_1 \) event is expressed as \( E^{\text{bit},X_1}_{ab} = E^{\text{bit},X_1}_{\Bar{a}\Bar{b}} \), where \( a \) and \( b \) (or \( \Bar{a} \) and \( \Bar{b} \)) represent the phase bits selected by Alice and Bob.

\subsubsection{Equal phase bits}
Here, we analyze the parameters of the \( X_1 \) event for the case where Alice's and Bob's phase bits are the same. We first calculate the gain \( Q^{H_1}_{00} \) as follows:
\begin{align}
Q^{H_1}_{00} = \frac{(1 - p_d)^3}{4} \hspace{-5mm}&\sum_{\text{X} \in \{++, +-, -+, --\}} \hspace{-5mm}
\exp\left(-\eta_d I_{\text{X}H_2{00}} - \eta_d I_{\text{X}V_1{00}} - \eta_d I_{\text{X}V_2{00}} \right) \nonumber\\
& \times\bigg[1 - (1 - p_d) \exp\left(-\eta_d I_{\text{X}H_1{00}}\right)\bigg].
\end{align}
The gain \( Q^{H_2}_{00} \) can be obtained similarly by swapping \( H_1 \) and \( H_2 \) in the corresponding detector settings.

The gain for the \( X_1 \) event in this case is then given by:

\begin{equation}
\begin{aligned}
Q^{X_1}_{00} &= Q^{H_1}_{00} + Q^{H_2}_{00}.
\label{QX_100}
\end{aligned}
\end{equation}
Following the earlier discussion, we now present the expression for the phase error rate \( E^{\text{ph},H_1}_{00} \).
\begin{align}
    E^{\text{ph},H_1}_{00} &= \frac{(1-p_d)^3}{4 Q^{H_1}_{00}} \sum_{\text{X}\in\{++, +-, -+, --\}} \hspace{-5mm}
    \exp\left(-I_{\text{X}H_1{00}} - I_{\text{X}H_2{00}} \eta_d \right. \nonumber\\
    &\quad \left. - I_{\text{X}V_1{00}} \eta_d - I_{\text{X}V_2{00}} \eta_d\right) \\
    &\times\bigg[\cosh(I_{\text{X}H_1{00}}) - (1 - p_d) \cosh\left((1 - \eta_d) I_{\text{X}H_1{00}}\right)\bigg]. \nonumber
\end{align}
The phase error rate \( E^{\text{ph},H_2}_{00} \) can be obtained similarly by swapping the corresponding indices \( H_1 \leftrightarrow H_2 \) in the expression for \( E^{\text{ph},H_1}_{00} \).
Consequently, the phase error rate of the \( X_1 \) event, when Alice and Bob choose equal phase bits, is written as
\begin{equation}
\begin{aligned}
    E^{\text{ph},X_1}_{00} &= \frac{Q^{H_1}_{00} E^{\text{ph},H_1}_{00} + Q^{H_2}_{00} E^{\text{ph},H_2}_{00}}{Q^{X_1}_{00}}.
    \label{EphX_100}
\end{aligned}
\end{equation}
Next, we present the expression for the bit error rate when Alice and Bob select equal phase bits in the \( X_1 \) event
\begin{equation}
\begin{aligned}
    E^{\text{bit},X_1}_{00} &= \frac{Q^{H_2}_{00}}{Q^{X_1}_{00}}.
\end{aligned}
\label{EbitX_100}
\end{equation}

\subsubsection{Different phase bits}
We now analyze the parameters of the \( X_1 \) event when Alice and Bob select different phase bits.
The gains \( Q^{H_1}_{01} \) and \( Q^{H_2}_{01} \) can be derived from \( Q^{H_1}_{00} \) and \( Q^{H_2}_{00} \), respectively, by replacing the subscript \( (00) \) with \( (01) \) in the corresponding intensities. To obtain the phase error rates \( E^{\text{ph},H_1}_{01} \) and \( E^{\text{ph},H_2}_{01} \), we apply the same process by replacing all instances of the subscript \( (00) \) with \( (01) \) in the expressions for \( E^{\text{ph},H_1}_{00} \) and \( E^{\text{ph},H_2}_{00} \). The phase error rate of the $X_1$ event for different phase bits chosen by Alice and Bob can be calculated in the same way as in Eq.~(\ref{EbitX_100}).

The bit error rate of the \( X_1 \) event for different phase bits chosen by Alice and Bob is expressed as:
\begin{equation}
\begin{aligned}
    E^{\text{bit},X_1}_{01} &= \frac{Q^{H_1}_{01}}{Q^{X_1}_{01}}.
\end{aligned}
\end{equation}

Finally, the overall gain, phase error rate, and bit error rate of the \( X_1 \) event can be expressed as follows:
\begin{equation}
\begin{aligned}
    Q^{X_1} &= \frac{1}{2} Q^{X_1}_{00} + \frac{1}{2}Q^{X_1}_{01}, \\
    E^{X_1}_\text{ph} &= \frac{1}{2} E^{\text{ph},X_1}_{00} + \frac{1}{2}E^{\text{ph},X_1}_{01}, \\
    E^{X_1}_\text{bit} &= \frac{1}{2} E^{\text{bit},X_1}_{00} + \frac{1}{2}E^{\text{bit},X_1}_{01}.
    \label{AllparamX_1}
\end{aligned}
\end{equation}

\subsection{\texorpdfstring{$X_2$}{X\_2} event}
The term \( Q^{H_{1(2)}V_{1(2)}}_{ab} = Q^{H_{1(2)}V_{1(2)}}_{\Bar{a}\Bar{b}} \) represents the gain of detectors \( H_{1(2)} \) and \( V_{1(2)} \) when Alice and Bob select phase bits \( a \) and \( b \) or \( \Bar{a} \) and \( \Bar{b} \). Similarly, the phase error rate \( E^{\text{ph},H_{1(2)}V_{1(2)}}_{ab} = E^{\text{ph},H_{1(2)}V_{1(2)}}_{\Bar{a}\Bar{b}} \) corresponds to the scenario when both detectors \( H_{1(2)} \) and \( V_{1(2)} \) click for the chosen phase bits. The bit error rate for the \( X2 \) event is given by \( E^{\text{bit},X2}_{ab} = E^{\text{bit},X2}_{\Bar{a}\Bar{b}} \), where \( a \) and \( b \) (or \( \Bar{a} \) and \( \Bar{b} \)) are the phase bits chosen by Alice and Bob.

\subsubsection{Equal phase bits}
We now analyze the parameters of the \( X2 \) event when Alice and Bob select identical phase bits, beginning with the calculation of the gain for detectors \( H_1 \) and \( V_1 \).
\begin{align}
    Q^{H_{1}V_{1}}_{00} &= \frac{(1 - p_d)^2}{2} \hspace{-2mm}\sum_{\text{X} \in \{++, +-, -+, --\}} \hspace{-5mm} 
    \exp\left(-\eta_dI_{\text{X}H_{2}{00}} -\eta_d I_{\text{X}V_{2}{00}}\right) \nonumber\\
    &\quad \times\bigg[1 - (1 - p_d) \exp\left(-\eta_d I_{\text{X}H_{1}{00}}\right)\bigg] \nonumber\\
    &\quad \times\bigg[1 - (1 - p_d) \exp\left(-\eta_d I_{\text{X}V_{1}{00}}\right)\bigg].
\end{align}
The gain \( Q^{H_2V_2}_{00} \) can be obtained from \( Q^{H_1V_1}_{00} \) by swapping \( H_1 \) with \( H_2 \) and \( V_1 \) with \( V_2 \).

Similar to Eq. (\ref{QX_100}), the overall gain of the \( X2 \) event, in the case where Alice's and Bob's phase bits are equal ($Q^{X2}_{00}$), is given by the sum of $Q^{H_2V_2}_{00}$ and $Q^{H_1V_1}_{00}$.
The phase error rate of the $X2$ event for equal phase bits chosen by Alice and Bob when detectors $H_1$ and $V_1$ click can be expressed as:
\begin{widetext}
\begin{equation}
\begin{split}
    E^{\text{ph},H_1V_1}_{00} = \frac{(1 - p_d)^2}{4 Q^{H_1V_1}_{00}} \sum_{\text{X} \in \{++, +-, -+, --\}} 
    \exp\left(-I_{\text{X}H_1{00}} - I_{\text{X}V_1{00}} - \eta_d I_{\text{X}H_2{00}} -\eta_d I_{\text{X}V_2{00}}\right) \\
    \times\Bigg[2 \left(\sinh(I_{\text{X}H_1{00}}) - (1 - p_d) \sinh\left(I_{\text{X}H_1{00}}(1 - \eta_d)\right)\right)
    \left(\cosh(I_{\text{X}V_1{00}}) - (1 - p_d) \cosh\left(I_{\text{X}V_1{00}}(1 - \eta_d)\right)\right) \\
    + \left(\sinh(I_{\text{X}V_1{00}}) - (1 - p_d) \sinh\left(I_{\text{X}V_1{00}}(1 - \eta_d)\right)\right)
    \left(\cosh(I_{\text{X}H_1{00}}) - (1 - p_d) \cosh\left(I_{\text{X}H_1{00}}(1 - \eta_d)\right)\right) \\
    + \left(\cosh(I_{\text{X}V_1{00}}) - (1 - p_d) \cosh\left(I_{\text{X}V_1{00}}(1 - \eta_d)\right)\right)
    \left(\cosh(I_{\text{X}H_1{00}}) - (1 - p_d) \cosh\left(I_{\text{X}H_1{00}}(1 - \eta_d)\right)\right)\Bigg].
\end{split}
\end{equation}
\end{widetext}
The expression for \( E^{\text{ph},H_2V_2}_{00} \) can be obtained directly from \( E^{\text{ph},H_1V_1}_{00} \) by swapping the detector indices \( H_1 {\leftrightarrow} H_2 \) and \( V_1 {\leftrightarrow} V_2 \) in the corresponding terms.
The phase error rate of the $X_2$ event, when Alice and Bob choose equal phase bits, can be calculated similarly to Eq.~(\ref{EbitX_100}).

The bit error rate of the \( X_2 \) event, in the same scenario, is written as follows
\begin{equation}
\begin{aligned}
    E^{\text{bit},X2}_{00} &= \frac{(1 - p_d)^2}{4 Q^{X_2}_{00}}
    \hspace{-2mm}\sum_{\text{X} \in \{++, +-, -+, --\}} \hspace{-5mm}
    \Big(1 + \delta_{\text{X}, +-} + \delta_{\text{X}, -+}\Big) \\
    &\quad \times\exp\Big(-\eta_dI_{\text{X}H_2{00}} -\eta_d I_{\text{X}V_2{00}}\Big) \\
    &\quad \times \Bigg[1 - (1 - p_d) \exp(-\eta_dI_{\text{X}H_1{00}} )\Bigg] \\
    &\quad \times \Bigg[1 - (1 - p_d) \exp(-\eta_dI_{\text{X}V_1{00}} )\Bigg].
\end{aligned}
\label{EbitX200}
\end{equation}
where $\delta$ represents the Kronecker delta function.

\subsubsection{Different phase bits}
We now analyze the situation where Alice's and Bob's phase bits differ.

To obtain \( Q^{H_1V_1}_{01} \) and \( Q^{H_2V_2}_{01} \), replace \(\{00\}\) with \(\{01\}\) in \( Q^{H_1V_1}_{00} \) and \( Q^{H_2V_2}_{00} \). The gain of the $X_2$ event is the sum of these terms. Similarly, \( E^{\text{ph},H_1V_1}_{01} \) and \( E^{\text{ph},H_2V_2}_{01} \) are derived by the same substitution in \( E^{\text{ph},H_1V_1}_{00} \) and \( E^{\text{ph},H_2V_2}_{00} \). The phase error rate can be calculated as in Eq.~(\ref{EbitX_100}), and the bit error rate as in Eq.~(\ref{EbitX200}) by replacing \(\{00\}\) with \(\{01\}\).

The overall parameters for the $X2$ event can be expressed similarly to Eq.~(\ref{AllparamX_1}).

\subsection{\texorpdfstring{$X_3$}{X\_3} event}
Using the same approach, we calculate the gains, phase error rates, and bit error rates for the $X_3$ event, and the overall parameters of this event will be obtained. 

We are now prepared to calculate the overall secret key rate for the INI-QKD protocol, which is given by
\begin{equation}
R = \sum_{i=1}^{3} R^{X_i} = \sum_{i=1}^{3} Q^{X_i} \left[1 - H(E^{X_i}_\text{ph}) - f H(E^{X_i}_\text{bit}) - I_E^U \right].
\end{equation}

\bibliography{tfqkd}

\end{document}